\documentstyle[editedbook,epsfig,psfig,epsf]{mq}

\def\cm2{cm$^2$ }
\def\se1{s$^{-1}$ }

\begin{opening}
\title{SS\,433: the second Wolf-Rayet X-ray binary ?}
\author{Y. Fuchs$^{1,2}$, L. Koch-Miramond $^1$ \& P. \'Abrah\'am$^3$}
\institute{$^1$ Service d'Astrophysique, CEA/Saclay, bat. 709, 91191 Gif sur Yvette cedex, France.\\
$^2$ Universit\'e Paris VII, UFR de Physique, 5 place Jussieu, 75005 Paris, France.\\
$^3$ Konkoly Observatory, P.O. box 67, 1525 Budapest, Hungary}
\end{opening}

\runningtitle{SS\,433: the second Wolf-Rayet X-ray binary ?}
\runningauthor{Fuchs, Koch-Miramond \& \'Abrah\'am}

\begin{document}
\vspace{-0.5cm}
\begin{abstract}
{\small We present mid-infrared spectrophotometric observations of SS\,433
with ISOPHOT. The He{\footnotesize I}+He{\footnotesize II} lines in
both spectra of SS\,433 and of the Wolf-Rayet star WR\,147, a
wind-colliding WN8+BO5 binary system, closely match. The 2.5-12 $\mu$m
continuum radiation is due to an expanding wind free-free emission in
an intermediate case between optically thick and optically thin
regimes.  The inferred mass loss rate evaluation gives 
\mbox{$\sim 10^{-4}$ M$_\odot$.yr$^{-1}$.} 
Our results are consistent with a Wolf-Rayet-like
companion to the compact object in SS\,433.  A similar study for
Cygnus\,X-3 confirms the Wolf-Rayet-like nature of its companion,
although with a later WN type than previously suggested.}
\end{abstract}

\section{The mysterious SS\,433 system}
	\hspace*{0.5cm} SS\,433 has been discovered as a source of
	strong H$\alpha$ lines, and as
	a variable radio and X-ray source \cite{margon84}. 
	It is an X-ray binary with a $\sim 13$ days period, producing 
	bipolar relativistic (0.26\,c) jets undergoing a precession movement
	in \mbox{$\sim$ 162.5 days,}
	so covering a cone with an opening angle of 19.8$^\circ$ 
	and an axis of $\sim 78.8^\circ$ with the line of sight. This
	precession has indeed been observed in radio, at arcsec
	\cite{hjell} and milliarcsec scales \cite{verm}.
	So SS\,433 was actually the first microquasar discovered.
	In spite of intensive studies, no line has been
	associated to the companion star yet. It is generally accepted but
	has never been proved that SS\,433 is a neutron star + massive
	star binary system.

\section{Mid-IR observations of SS\,433}
   \subsection{Comparison with Wolf-Rayet stars spectra}
\begin{figure}[!htb]
	\psfig{file=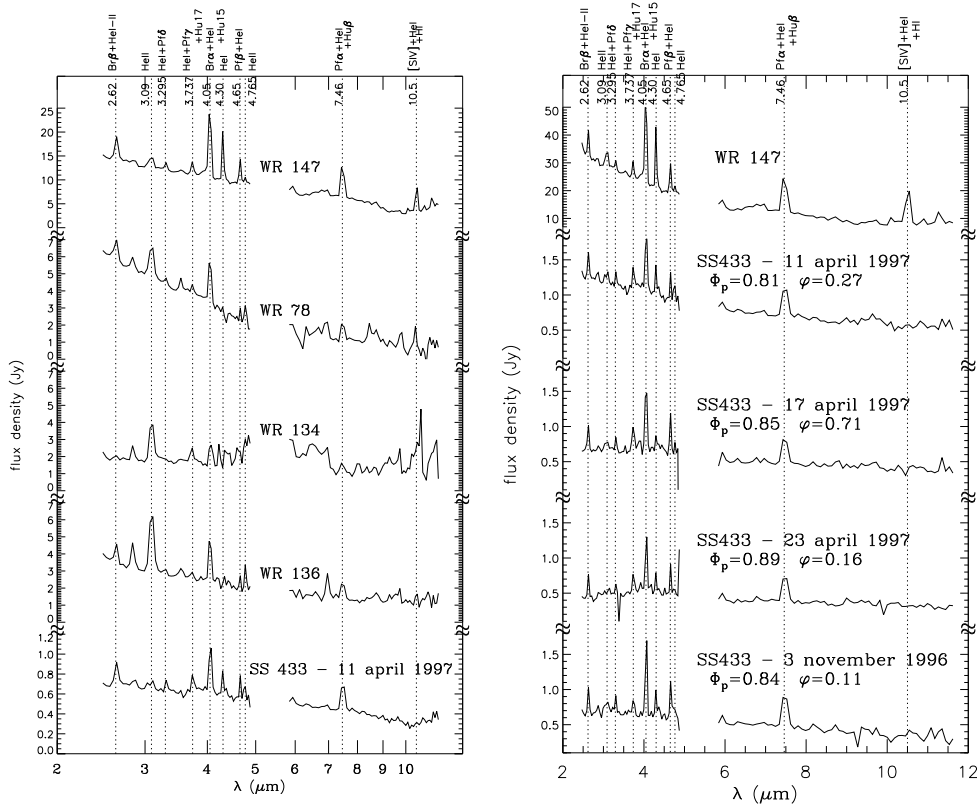,width=\linewidth}
	\vspace*{-0.8cm}
	\caption{{\it Left:} Comparison of SS\,433 observed spectra
	with that of Wolf-Rayet stars (of WN type). {\it Right:}
	Comparison of WR\,147 dereddened spectrum with the ones of SS\,433
	taken at different precession phases
	($\Phi_{\mathrm{p}}$) and orbital phases ($\varphi$). Lines
	are identified according to [5].}
	\label{figwr}
	\end{figure}

	\vspace*{-0.2cm}
   \hspace*{0.5cm} SS\,433 has been observed with ISOPHOT, in the
   spectral mode at 2.5-5 $\mu$m ($\sim$\,0.04\,$\mu$m resolution) 
   and 6-12 $\mu$m ($\sim$\,0.1\,$\mu$m resolution), and in the
   photometric mode at 12, 25 and 60 $\mu$m, in November 1996 and
   April 1997. We looked for Wolf-Rayet (WR) star observations in the
   ISO archives, as it was the suspected type for SS\,433 companion
   star. Several WRs were observed with ISOSWS, and so we rebined the
   SWS spectra to the spectral resolution of ISOPHOT.\\
	\hspace*{0.5cm} As a lot of hydrogen is seen in SS\,433
	optical spectra, we chose to compare it to WRs of nitrogen type (WN)
	which are in the first stage of WR
	evolution. Figure~\ref{figwr} (left) shows the observed spectra of the
	four corresponding WRs, classified from the not very evolved
	(late type WNL) to the evolved (early type WNE) ones: WR\,147
	$\leftrightarrow$ WN8+B0.5 (WNL), WR\,78 $\leftrightarrow$
	WN7h (WNL), WR\,136 $\leftrightarrow$ WN6b (WNE) and WR\,134
	$\leftrightarrow$ WN6 (WNE). Comparing on figure~1 these
	spectra to the one of SS\,433, it is clear that SS\,433 is closest to
	WR\,147, a late type WR.
   \newpage
	\vspace*{-1.2cm}

	\vspace*{-0.3cm}
	Figure~\ref{figwr} (right) also shows the
	comparison of WR\,147 spectrum with that of SS\,433 taken at
	different phases. SS\,433 and WR\,147 spectra were dereddened
	with A$_{\mathrm{V}}$ = 8 and 11.2 mag respectively, using the Lutz
	et al. \cite{lutz} law. The lines, identified according 
	to\,\cite{morris}, are strong and weak H+He{\footnotesize I}
	blended lines (because of ISOPHOT spectral resolution) and
	weak He{\footnotesize II} lines. No metallic line was
	observed. The SS\,433 line spectrum is clearly WR-like. We note
	that WR\,147 is known as a colliding wind binary system which
	may explain its resemblance to SS\,433.\\

	\vspace*{-0.7cm} 
   \subsection{Continuum} 
\begin{figure}[!ht]
	\centering
	\psfig{file=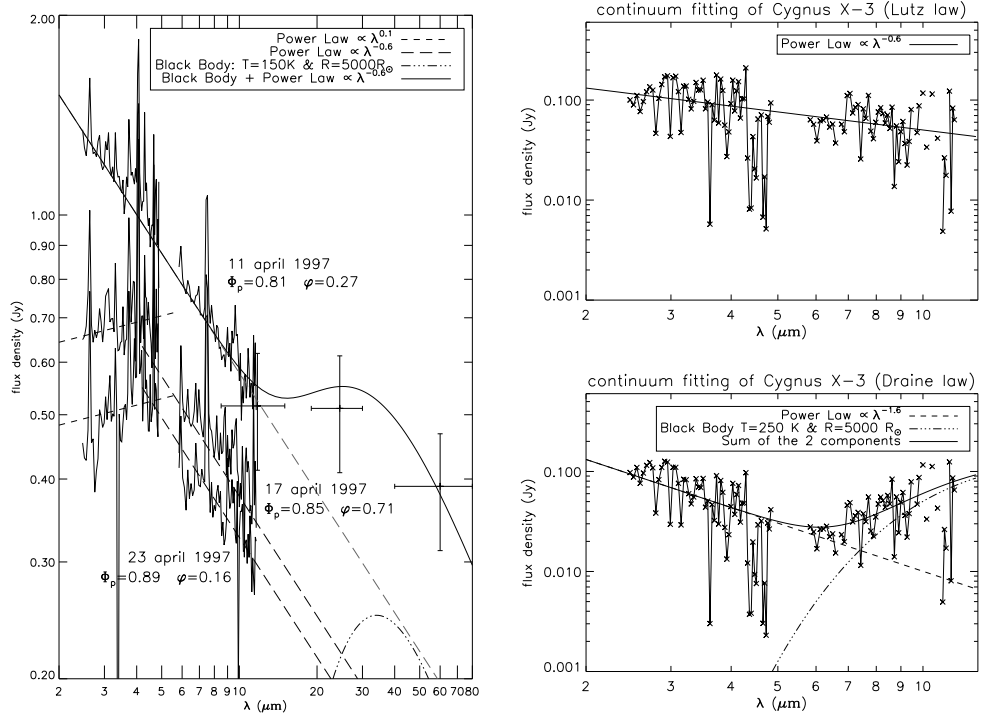,width=\linewidth}
	\vspace*{-0.8cm}
	\caption{{\it Left:} Continuum spectra of SS\,433. {\it Right:} Continuum spectra of Cygnus\,X-3.}
	\label{figcont}
	\end{figure}

	\vspace*{-0.2cm}
	\hspace*{0.5cm} Figure~\ref{figcont} shows the continuum
	spectra of SS\,433, which flux density (F$_\nu$) is well
	fitted by power laws in the 2-12 $\mu$m range, corresponding
	to free-free emission: optically thin for F$_\nu \propto
	\lambda^{0.1}$, and in the intermediate regime between
	optically thin and thick for F$_\nu \propto \lambda^{-0.6}$.
	In far-IR, an additional component, as a black-body
	due to dust surrounding the system, is needed. The 0.6
	spectral index is characteristic of an ionized homogeneous wind
	with a spherical expansion at a constant velocity \cite{wri},
	although it is also valid for a more complex geometry as long
	as it stays thick \cite{schmid}. So SS\,433 continuum spectrum
	corresponds to the one emitted by a standard O or WR wind.\\

	\vspace*{-0.7cm}
   \subsection{Mass loss rate}
	\vspace*{-0.2cm}
	\hspace*{0.5cm} We calculated the mass loss rate of this
	free-free emitting wind, following the Wright \& Barlow\,\cite{wri} 
	formula\,(8). With a distance D\,=\,3.5\,kpc, a Gaunt
	factor g\,$\sim$\,1, a \mbox{F$_\nu$\,=\,1000\,mJy} flux at
	4\,$\mu$m ($7.5 \times 10^{13}$\,Hz), and for a WN-type wind
	where the mean atomic weight per nucleon $\mu\,=\,1.5$, the
	number of free electrons per nucleon $\gamma\,=\,1$, the mean
	ionic charge Z\,=\,1 and the velocity 
	$v_\infty\,=\,1000$\,km.s$^{-1}$, we find:
	\.M $= 1.0 \times 10^{-4}$ M$_\odot$.yr$^{-1}$.
	However, the recent WN mass-loss rate estimates
	\cite{morris99} show that this value has to be lowered by a
	factor 2 or 3 due to clumping in the wind.
        This is in good agreement with the mass transfer rate
        estimated by van den Heuvel et al.\,\cite{heuvel} assuming a normal
        homogeneous WR wind, or with the recent mass transfer values
        obtained from simulations of SS\,433 evolution by King et
        al.\,\cite{king}. 
	For WR\,147, \.M $\simeq 1.5-3.7 \times 10^{-5}$
	M$_\odot$.yr$^{-1}$ and the WN mean is \mbox{$\sim 3 \times
	10^{-5}$ M$_\odot$.yr$^{-1}$,} so our $\sim 10^{-4}$
	M$_\odot$.yr$^{-1}$ mass loss evaluation for SS\,433 is
	compatible with a strong WNL wind.\\
	\vspace*{-0.8cm}

	\vspace*{-0.3cm}
	   What happens to the material contained in this wind is
	   unknown since only \mbox{$\sim$\,$10^{-7}$\,M$_\odot$.yr$^{-1}$} is
	   ejected into the jets \cite{marshall}. It may be accreted
	   then ejected via the L$_2$ point \cite{fabrika}, or close
	   to the compact object \cite{king}. It may then form dust
	   seen in far-IR at a distance $> 20$\,a.u. This material is
	   probably responsible for the equatorial outflows observed
	   with the VLBA \cite{paragi}\cite{blund} at $\sim
	   100$\,a.u. from SS\,433.
	
\section{Cygnus\,X-3}
\begin{figure}[!htb]
	\centering
	\vspace*{0.4cm}	\psfig{file=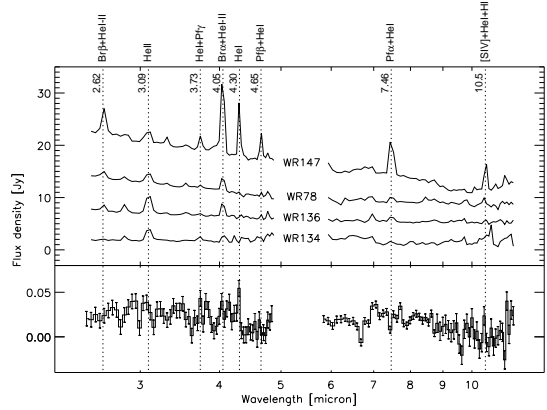,width=6.5cm}
	\vspace*{0.2cm}\caption{Observed spectra of Cygnus\,X-3 and four WR stars. An arbitrary vertical offset has been added to the WR spectra for clarity.}
	\label{figcygx3}
	\vspace*{-0.2cm}	
	\end{figure}

   \hspace*{0.5cm} Cygnus\,X-3 is a microquasar for which WR-like
   features have been detected in near-IR spectra \cite{vanker}\cite{fender}. 
   We also observed it
   with ISOPHOT-S during its quiescent state, but with much more
   difficulties as it is a weak IR source. Its spectrum was also
   compared to the four WR spectra as shown in
   figure~\ref{figcygx3}. Only one line was found in Cygnus\,X-3
   spectrum, at 4.3\,$\mu$m with a confidence level of more than
   4.3\,$\sigma$, corresponding to an He{\footnotesize I} line only
   visible in WR\,147 spectrum.\\

	\vspace*{-0.8cm}
   The spectral fitting of Cygnus\,X-3 dereddened 
   (A$_{\mathrm{V}}$\,=\,20\,mag)
   spectra is obtained with a unique power law
   F$_\nu$\,$\propto$\,$\lambda$$^{-0.6}$ in the 2.4-12\,$\mu$m range
   when dereddened with the Lutz et al.\,\cite{lutz} law
   (figure\,\ref{figcont} upper right), and with the sum of a power law
   with slope $\lambda$$^{-1.6}$ and a black body, a hint for the
   presence of circumstellar dust, when using the Draine law\,\cite{draine} 
   (figure\,\ref{figcont} lower right). We cannot choose
   between these two laws since the molecular composition of the
   absorbing material on the line of sight to Cygnus\,X-3 is
   unknown. As for SS\,433, the power law part of the continuum
   spectrum can be explained by free-free emission of an expanding wind
   in the intermediate case between optically thick and optically
   thin regimes. The corresponding mass loss evaluation, with 
   63\,mJy at 6.75\,$\mu$m and $v_\infty$=1500\,km.s$^{-1}$\cite{vanker}, 
   gives \.M $=1.2\times 10^{-4}$\,M$_\odot$.yr$^{-1}$.
   We conclude \cite{koch} that our results are
   consistent with a Wolf-Rayet companion to the
   compact object in Cygnus\,X-3 of WN8 type, a later type than suggested by
   previous works \cite{vanker}\cite{fender}. 

\section{Conclusion}
\vspace*{-0.2cm}
	\hspace*{0.5cm} SS\,433 shows a WN8-like spectrum, which
	continuum is due to free-free emission corresponding to a
	standard O or WR wind or a geometrically complex thick
	wind. The inferred mass loss of $\sim 10^{-4}$
	M$_\odot$.yr$^{-1}$ is compatible with a WNL wind. So either
	SS\,433 is the second Wolf-Rayet X-ray binary known in our
	Galaxy after Cygnus X-3, or it shows WR-like conditions, and
	at that point we cannot distinguish between these two cases.

\end{document}